# Impedance matching and emission properties of optical antennas in a nanophotonic circuit

Jer-Shing Huang, Thorsten Feichtner, Paolo Biagioni & Bert Hecht[*]

*Nano-Optics & Biophotonics group, Department of Experimental Physics 5, Röntgen Research Center for Complex Material Research (RCCM), Physics Institute, University of Würzburg, Am Hubland, D-97074 Würzburg, Germany.*

---

[*] Email: hecht@physik.uni-wuerzburg.de

An experimentally realizable prototype nanophotonic circuit consisting of a receiving and an emitting nano antenna connected by a two-wire optical transmission line is studied using finite-difference time- and frequency-domain simulations. To optimize the coupling between nanophotonic circuit elements we apply impedance matching concepts in analogy to radio frequency technology. We show that the degree of impedance matching, and in particular the impedance of the transmitting nano antenna, can be inferred from the experimentally accessible standing wave pattern on the transmission line. We demonstrate the possibility of matching the nano antenna impedance to the transmission line characteristic impedance by variations of the antenna length and width realizable by modern microfabrication techniques. The radiation efficiency of the transmitting antenna also depends on its geometry but is independent of the degree of impedance matching. Our systems approach to nanophotonics provides the basis for realizing general nanophotonic circuits and a large variety of derived novel devices.

**Introduction**

Miniaturization and packaging density of integrated optics based on dielectrics is limited by the wavelength scale modal profiles of guided modes.[1] In contrast, plasmonic modes on noble metal nanostructures offer strong subwavelength confinement and therefore promise the realization of nanometer-scale integrated optical circuitry.[2,3] A truly subwavelength integrated photonic circuit based on plasmonic nano structures will generally consist of (i) a set of optical antennas[4,5,6,7] to efficiently excite specific local modes by far-field radiation, (ii) a very small footprint network of optical transmission lines[8] (OTLs) to distribute and manipulate plasmonic excitations,[9,10,11,12,13,14,15,16] and (iii) another set of optical antennas to efficiently convert local modes into propagating photons. The properties of metal nanoparticle chains,[17,18] metal nanowires,[19,20] line defects in plasmonic photonic crystals,[21] as well as gaps[22,23,24,] and v-shaped grooves[9,25,26] in extended metal films have been explored as subwavelength waveguides for light. Efficient launching of specific guided modes on such structures is difficult since it requires matching of both, the small mode extension and the k-vector. It has been shown recently that efficient coupling between far-field photons and subwavelength spatial domains can be achieved using resonant optical antennas.[5,7,27,28,29,30,31] However, so far optical antennas have mostly been studied as isolated elements. Here we consider optical antennas as integral parts of an experimentally realizable integrated nanophotonic circuit where they act as efficient interfacing elements between propagating photons and guided modes of a plasmonic two-wire transmission line. We show by simulations that the principles of classical transmission line theory, e.g. impedance matching,[32] between the two-wire OTL and dipole antennas are fully applicable at optical frequencies. We further suggest that complex impedances of circuit elements at optical wavelength[33,34] may be experimentally obtained in terms of reflection coefficients. We finally determine the radiation efficiency of the emitting optical antenna as well as the overall efficiency of the nanophotonic circuit.

**Methods, structural parameters and numerical simulation**

The system simulated by the finite-difference time-domain method (FDTD Solutions, version 5.1.2, Lumerical Solutions Inc., Vancouver, Canada) contains two dipole nano antennas connected by a nanosize two-wire OTL with a 30×30 nm$^2$ quadratic wire cross section and a wire separation of 10 nm equal to the feed-gap width of both antennas (see Fig. 1). The structure is made of gold ($\varepsilon$ = -26.57+1.66j)[35] and situated in air ($\varepsilon$ = 1) on top of a glass half space ($\varepsilon$ = 2.11). One of the antennas (left side, receiving antenna) is illuminated by a tightly focused Gaussian beam (spot size=340 nm, $\lambda$=830 nm) from within the glass half space with polarization parallel to the antenna. Enhanced and confined optical fields created in the antenna feed-gap are efficiently converted into guided transverse electric (TE) modes propagating along the transmission line, which are finally converted into far-field photons by the second antenna (emitting antenna). The equivalent circuit of the system is shown in the inset of Fig. 1. It consists of a generator providing an ac-voltage at the chosen optical frequency including the generator impedance $Z_G$ representing the receiving antenna, a transmission line with characteristic impedance $Z_0$, as well as a load impedance $Z_L$ representing the emitting antenna. In the following we show that the coupling efficiency between elements of this prototype nanophotonic circuit is governed by the degree of impedance matching. We also outline how impedance matching can be achieved by variation of experimentally accessible geometrical parameters.

**Optimization of the receiving nano antenna**

As a first step we optimize the coupling of the receiving antenna to an infinitely long OTL, which is characterized by its characteristic impedance, $Z_0$.[32] To mimic an infinitely long transmission line in the simulation, the OTL extends long enough (>3000 nm) inside the simulation area before being terminated by the perfect match layer boundary. Due to the considerable damping of the guided modes on the OTL the amplitude of reflected fields reaching the receiving antenna is negligible, as required for a system of infinite length. In order to optimize the coupling of far-field power into the OTL guided modes we vary the length of the receiving antenna (cross section 30×30 nm$^2$) and record the intensity enhancement in the antenna feed-gap (point P) and at an equivalent position at a distance of 1000 nm down the OTL (point Q). The results are plotted in Fig. 2A along with the intensity enhancement obtained in the feed-gap of an isolated dipole antenna of the same geometry. Compared to the isolated dipole antenna, the OTL-connected antenna shows a lower intensity enhancement but exhibits a resonance at about the same total length of 200 nm, which is kept constant in all further simulations. We tentatively conclude that attaching a two-wire OTL oriented perpendicular to the electric fields in the feed-gap does not significantly affect the antenna resonance length. It is also evident that the maxima of the intensity enhancements at point P and Q coincide. This implies that the intensity of the OTL guided modes is proportional to the intensity enhancement in the feed-gap of the receiving antenna. Compared to the same transmission line without receiving antenna, the far-field power coupling improves by more than a factor of 200 with a resonant receiving antenna.

**Properties of the nanosize two-wire optical transmission line**

Having efficiently launched guided modes on the OTL we are interested in their propagation constants.[32] Fig. 2B shows the intensity enhancement along a line marking the center of the transmission line (red cross in the inset), while Fig. 2C shows the intensity enhancement in a plane 15 nm above and parallel to the glass/air interface (red dotted line in the inset). A strong exponential damping of the intensity enhancement is observed with superimposed slow undulations close to the receiving antenna. The slow undulations are due to the beating of at least three modes with different propagation constants (see supplementary material, Fig. S1). After a distance of 500 nm from the receiving antenna these undulations cease and only the fundamental mode prevails. A very small modulation is observed close to the end of the simulation area due to a weak reflection from the perfectly-matched-layer boundary. Exploiting this modulation and the overall exponential decay we determine the complex propagation constant of the guided fundamental TE mode $\gamma = (0.00084+0.02802j)$ nm$^{-1}$ by nonlinear fitting of the curve in Fig. 2B. From this we obtain a decay length of 1190 nm and an effective wavelength of 224 nm, which is much shorter than the free space wavelength.[36,37] An independent simulation using the full-vectorial finite-difference frequency-domain (FDFD) method[38] (MODE Solutions, version 3.0.1, Lumerical Solutions Inc., Vancouver, Canada) yields similar results. The propagation constants of the two higher-order modes (one very short-ranged, one longer-ranged but very low intensity) are determined by a stepwise fitting of respective segments of the decay (see supplementary material, Fig. S1). The inset of Fig. 2B shows the near-field intensity modal profile of the fundamental mode. The corresponding distributions of the electric (**E**) and magnetic (**H**) field components are plotted in Fig. S2 (supplementary material). The characteristic impedance of the two-wire OTL is defined as $Z_0 = V/I$. The voltage *V* between two finite-width wires is obtained by a line integral over the complex electric field **E** from one wire core to the other, as indicated by the red-dashed line in Fig. S2, while the current *I* is evaluated based on Ampère's law. Since the magnetic field **H** is zero at infinity, we replace the typical close-loop integration path with a sufficiently long path approximating a linear path from $y = -\infty$ to $y = +\infty$. The OTL characteristic impedance thus obtained has a value of $Z_0 = (216-5.5j)$ Ω, which is comparable to impedances of two-wire transmission lines at radio frequencies.[32] The robustness of this method is confirmed by the very weak dependence of $Z_0$ on the choice of the integration path ($\Delta Z_0 < 6\%$), which is due to the weak inhomogeneity of the field distribution inside the gap.

**Impedance matching in a complete nanophotonic circuit**

Knowing the complex propagation constant of the fundamental TE mode supported by the two-wire OTL and its characteristic impedance we are able to quantitatively study effects that occur in nanophotonic circuits based on finite-length OTLs. In Fig. 3, we compare a finite-length open OTL (*L*=1344 nm, Fig. 3A) with the same OTL terminated by an emitting antenna serving as an arbitrary load (Fig. 3B). Corresponding 2D images of the intensity enhancement recorded in the same plane as Fig. 1C are shown in the lower panel of Fig. 3. Due to a finite impedance mismatch between the OTL and the load we observe characteristic standing wave patterns. The complex voltage reflection coefficient $\Gamma_V$ for the discontinuity can be determined by nonlinear fitting of these standing wave patterns between the centers of two antennas. As to be expected from classical transmission line theory, the open end termination of the OTL leads to a near 100% reflectivity ($|\Gamma_V| = 95.6\%$) while the termination with the particular antenna of Fig. 3B results in a much lower reflectivity ($|\Gamma_V| = 37.3\%$). The deviation from perfect reflection of the open ended OTL is due to residual radiation from the truncated finite-width nanowires. Since the reflectivity is a direct measure of the impedance matching between the OTL and the load, the load impedance $Z_L$ can be determined according to

$$Z_L = Z_0 \frac{1+\Gamma_V}{1-\Gamma_V} \qquad (1)$$

once the complex reflection coefficient is known.[32] This approach avoids ambiguities that occur due to the fact that at optical frequencies, as opposed to the radiowave and microwave regime, the finite extension of nanostructures in the nanophotonic circuit can no longer be neglected since the fields vary strongly over the emitting antenna's gap volume (see for example Fig. 3B). Therefore, the concept of a feed "point" is difficult to apply, which makes it impractical to evaluate impedances by calculating the voltage to current ratio. From an experimental point of view, since plasmon propagation and standing waves on the OTL can be observed experimentally, e.g. by photoemission electron microscopy[39,40] or scanning near-field optical microscopy,[41,42] it is possible to use this approach to determine impedances of circuit elements at optical frequencies.

To minimize the voltage reflection coefficient at the load antenna, we scan the length and the width of the emitting antenna to achieve optimal impedance matching. While the effects of scanning the antenna length can be understood in analogy to classical antenna theory, we assume that by changing the antenna width we change the effective wavelength of the plasmonic mode on the nanowire.[36] Fig. 4A shows the voltage reflectivity ($|\Gamma_V| = \sqrt{\Gamma_V \Gamma_V^*}$) as a function of the total length of the emitting antenna varied between 70 nm and 610 nm. Plotted in different colors are traces obtained for antennas with widths of 30, 40, 50, and 60 nm, while the structure height is always kept constant at 30 nm. All plots clearly show two distinct minima, which can be related to the respective best matching of the antenna impedance to $Z_0$ for a given width. Using (1), the input impedances of the respective emitting antennas can be calculated. The results are plotted in Fig. 4B. For a fixed antenna width, the antenna impedance describes a spiral in the complex impedance plane where the position of $Z_0$ is also marked. We observe that as the antenna width is increased the average radius of the impedance spirals decreases while simultaneously the spirals shift to smaller resistances. This behaviour is in full analogy to that of radiowave antennas.[43] A shorter distance $|Z_L - Z_0|$ in the complex $Z_L$ plane indicates better impedance matching. The shortest distance $|Z_L - Z_0|$ on the plane (Fig. 4B) obtained for a 610 nm long and 50 nm wide antenna corresponds to the minimum reflectivity ($|\Gamma_V| = 2.7\%$) in Fig. 4A.

We note that there are other parameters, apart from the antenna width, that one may change to optimize the impedance matching between two-wire OTL and antenna, e.g. the widths of the OTL's wires and the gap between them, as well as the connection geometry.[43] Furthermore, the analogy to classical transmission line theory suggests that passive elements such as "stubs" may be used in passive impedance matching circuits[32] at optical frequencies.

**Power reflection and total efficiency**

Having optimized the impedance matching between the OTL and the emitting antenna, we are interested in the radiation efficiency of the emitting antenna as well as in the overall power transmission of the circuit. So far we have shown that by changing structural dimensions of the involved circuit elements, impedance matching can be achieved and $|\Gamma_V|$ can be minimized. The minimum $|\Gamma_V|$, however, only results in the maximum transportation of power if $Z_0$ is real. Since $Z_0$ is complex in our work, the time-averaged power delivered to the load is not given by the difference between the powers of the incident and reflected wave.[44] We therefore employ the concept of power waves and Kurokawa's method[45] to describe the power flow in the nanophotonic circuit. Using the concept of power waves, the power reflection coefficient $\Gamma_P$ can be expressed as

$$\Gamma_P = \left| \frac{Z_L - Z_0^*}{Z_L + Z_0} \right|^2 \tag{2}$$

where $Z_0$ represents the impedance seen by the emitting antenna (load). $\Gamma_P$ calculated according to (2) is plotted in Fig. 5A while an equivalent Smith chart, mapping the modified impedance as mentioned by Kurokawa[45] and used in passive RFID design,[46] is presented in Fig. S3. Compared to the voltage reflection coefficient $\Gamma_V$ in Fig. 4A, $\Gamma_P$ shows a similar trend with small deviations due to the small imaginary part of $Z_0$. For the design of structures with high power transmission efficiency, the modified Smith chart (Fig. S3) provides a quick guideline by considering the distance of the respective load impedance to the chart center, which is a direct measure of $\Gamma_P$. Since $\Gamma_P$ describes the power reflection, a minimal $\Gamma_P$ should result in a maximum power input to the respective emitting antenna. We determine the net time-averaged power flow $P_{in}$ along the OTL, 95 nm in front of the emitting antenna, by integrating the Poynting vector over a respective plane perpendicular to the OTL, and normalize it with the injected source power $P_{source}$. As shown in Fig. 5B, the maxima of $P_{in}/P_{source}$ coincide with the minima of the power reflection coefficient. To evaluate the radiation efficiency of the emitting antenna, we obtain the power $P_{out}$ radiated from the emitting antenna by integrating the outward Poynting vector over a closed box that encloses the emitting antenna but excludes a small rectangular area where the transmission line enters the box. We calculate the radiation efficiency $\eta_{out}$ of the emitting antenna as[43]

$$\eta_{out} = \frac{P_{in}}{P_{out}}. \tag{3}$$

Fig 5C shows the radiation efficiency as a function of the total antenna length for antenna widths varying from 30 nm to 60 nm. For all antennas the radiation efficiencies increase markedly from close to 0 to about 60% when the antenna length is increased to 200 nm. The first maxima of the radiation efficiency curves seem to coincide with the first minima of the reflectivity curves in Fig. 5A, a behavior which is not observed for the second reflectivity minima. Indeed, we find that the concomitance of best conditions for load impedance matching and for maximum radiation efficiency for the first reflectivity minimum is a coincidence. As may be inferred from the deviations between reflectivity minima and radiation efficiency maxima for the second resonances there is in general no reason why an antenna that provides good impedance matching should also provide high radiation efficiency.

To estimate the overall performance of the nanophotonic circuit, we finally consider the power flow through the system for the case of best matching, i.e. for a 230 nm long and 50 nm wide emitting antenna. The time-averaged power radiated from the emitting antenna $P_{out}$ can be expressed in terms of the time-averaged Gaussian source power $P_{source}$ as

$$P_{out} = \eta_{tot} \times P_{source} = \eta_{out} \times (1-\Gamma_P) \times e^{-2\alpha L} \times \eta_{in} \times P_{source} \qquad (3)$$

where $\eta_{tot}$ is the total efficiency and $\eta_{in}$ is the ratio between the time-averaged power coupled into the OTL by the receiving antenna and the source power $P_{source}$. The factors $(1-\Gamma_P)$ and $e^{-2\alpha L}$, respectively, take into account the power reflection at the emitting antenna and losses along the OTL. We obtain a total efficiency $\eta_{tot} = 2.65\%$ and an efficiency of the receiving antenna of $\eta_{in} = 27.6\%$. The fairly large value for $\eta_{in}$ provides quantitative evidence that the receiving antenna is an efficient collector of far-field radiation, while the much lower value for the overall efficiency $\eta_{tot}$, is dominated by propagation losses.

**Conclusions**

In conclusion, we have proposed and analyzed an experimentally realizable prototype nanophotonic circuit consisting of a receiving and emitting optical antennas connected by a nanosize two-wire OTL. By evaluating standing wave patterns we are able to investigate impedance matching and absolute impedances of circuit elements at optical frequencies. We also point out that a well-matched antenna does not necessarily have to have the highest radiation efficiency. The evaluation of standing wave patterns should be applicable in practical realizations of nanophotonic circuits to experimentally determine impedances of nanophotonic circuit elements. Finally, we analyze the power flow and give a quantitative estimation of the relevant efficiencies for power transfer in the system. Our work clearly shows that systems of interconnected nanophotonic elements can be optimized by applying concepts of impedance matching which may also be applied to understand and control the coupling between single quantum emitters and metallic nanostructures.

**Acknowledgements**

The authors would like to thank L. Su, C. S. Huang, C. Capsoni, M. D'Amico, G. Gentili, and J. Kern for useful scientific discussions. P. Biagioni acknowledges a Humboldt Research Fellowship for a Postdoctoral Researcher.

**Figure and Captions**

**Fig. 1**: Huang *et al.*

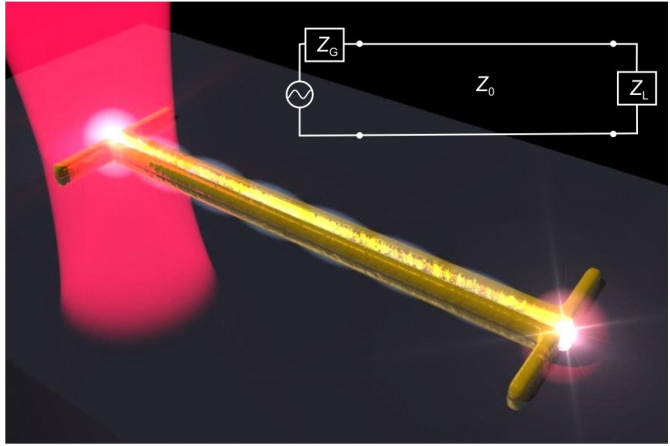

**Fig. 1**: Artists view of the overall geometry of a nanophotonic circuit consisting of a receiving antenna (left), a two-wire optical transmission line as well as an emitting antenna. The receiving antenna is excited by a linear polarized Gaussian source and launches guided modes which propagate along the nanosize two-wire optical transmission line. The emitting antenna (right) converts guided modes into propagating photons. The inset shows the corresponding equivalent circuit consisting of a generator, a transmission line, and a load.

**Fig. 2**: Huang *et al.*

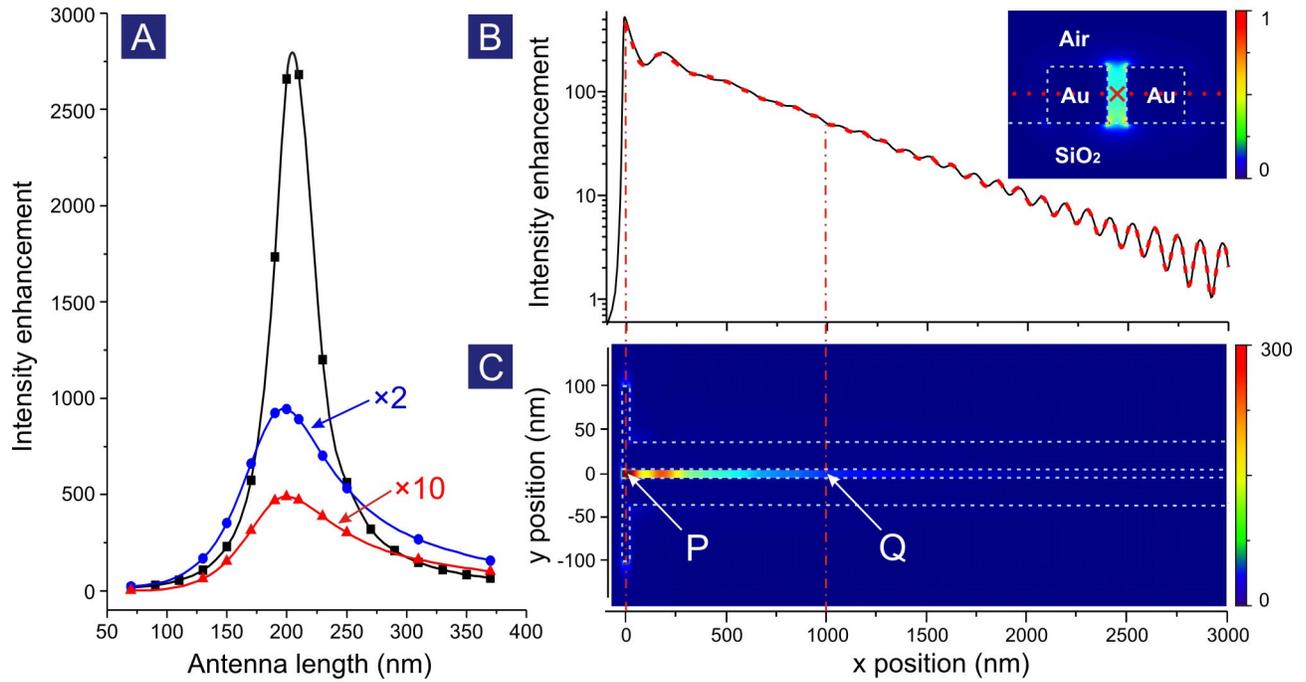

**Fig. 2**: Receiving antenna and infinitely long optical transmission line. (A) Intensity enhancement in the feed-gap as a function of the total antenna length for an isolated dipole antenna (black squares), the gap of an antenna connecting to the optical transmission line (multiplication factor ×2, blue dots), as well as the field enhancement at an equivalent position 1000 nm down the optical transmission line (multiplication factor ×10, red triangles). (B) Line cut of the intensity enhancement along the center of the transmission line. Red dashed line shows the best fit. The inset shows the normalized modal intensity profile of the fundamental guided TE mode. (C) 2D map of the intensity enhancement recorded in a plane parallel to the substrate, 15 nm above the glass/air interface, indicated by the red dotted line in the inset of (B). Note the unequal scales in x and y direction.

**Fig. 3**: Huang *et al*.

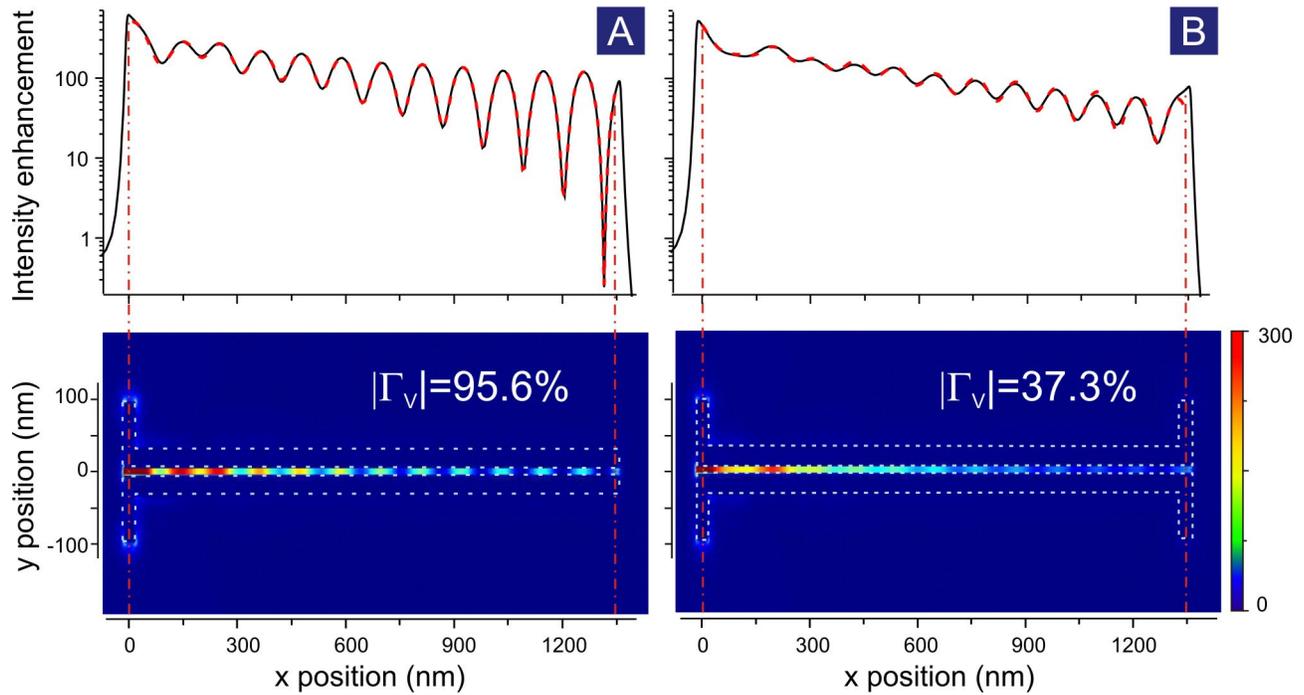

**Fig. 3**: Effect of different loads on reflectivity. Upper panel: Line cut and best fit (red dashed line) of the intensity enhancement in the gap of a 1344 nm long optical transmission line terminated by (A) an open end (antenna length 70 nm) and (B) a 200 nm long emitting antenna. Lower panel: corresponding 2D field enhancement maps. The respective voltage reflectivities at the terminations are indicated. Note the unequal scales in x and y direction.

**Fig. 4**: Huang *et al.*

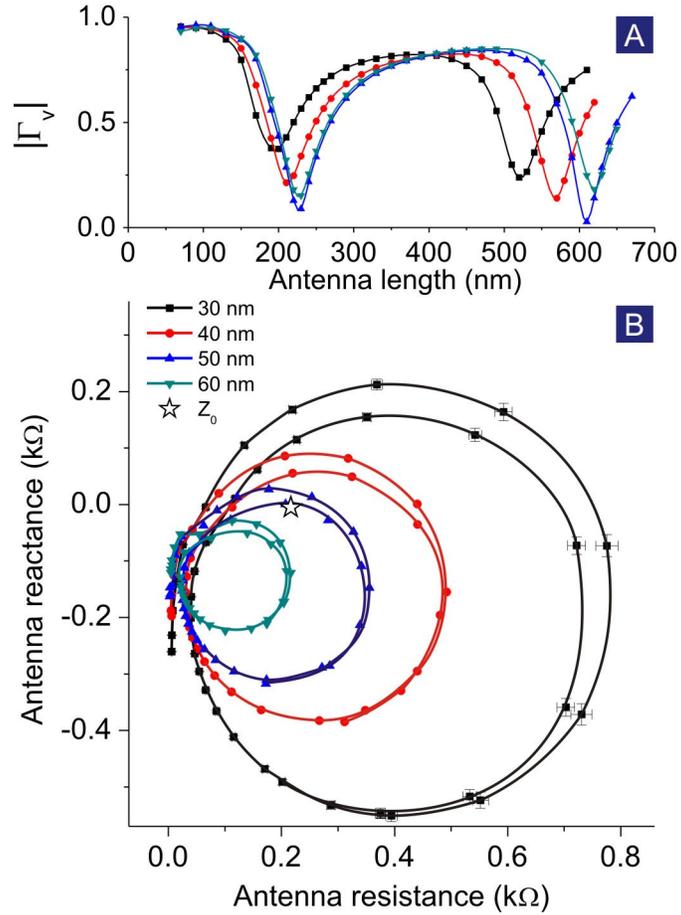

**Fig. 4**: Emitting antenna: voltage reflectivity and antenna impedance. (A) Voltage reflectivity ($|\Gamma_V| = \sqrt{\Gamma_V \Gamma_V^*}$) as a function of emitting antenna's total length for antenna widths of 30 nm (black), 40 nm (red), 50 nm (blue), and 60 nm (green). (B) Input impedance of the emitting antenna in the complex $Z_L$ plane. Representative error bars resulting from uncertainties of the reflectivity's amplitude and phase are displayed for the 30 nm wide antenna. The open star represents the position of $Z_0$. Black arrows and circles denote the point of best impedance matching corresponding to the closest approach of antenna impedance and $Z_0$. The grey arrows indicate the direction of increasing total antenna length.

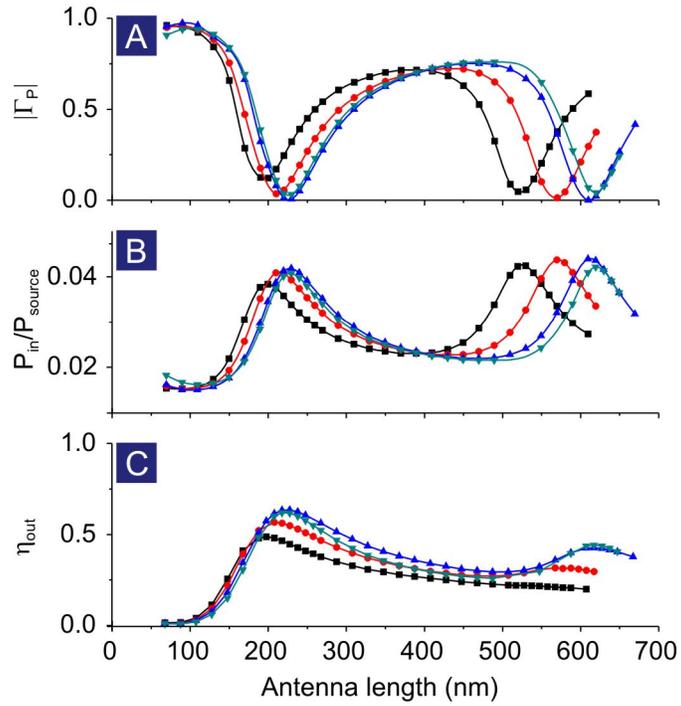

**Fig. 5**: Huang *et al*.

**Fig. 5**: Emitting antenna: power delivered and radiation efficiency. (A) Power reflection coefficient as a function of the antenna length for antenna widths of 30 nm (black), 40 nm (red), 50 nm (blue) and 60 nm (green). (B) input power and (C) radiation efficiency of the emitting antenna, respectively, for the corresponding antenna dimensions. The non-zero offset of $P_{in}$ is due to the finite radiation of the open-ended OTL and losses in the material.

# References

[1] Yariv, A. *Optical Electronics in Modern Communications* (Oxford University Press, Oxford, 1997)

[2] Barnes, W. L., Dereux, A. & Ebbesen, T. W. Surface plasmon subwavelength optics. *Nature* **424**, 824 (2003).

[3] Ozbay, E. Plasmonics: merging photonics and electronics at nanoscale dimensions. *Science* **311**, 189 (2006).

[4] Pohl, D. W. Near field optics seen as an antenna problem. *Near field optics: principles and applications / The second Asia-Pacific workshop on near field optics* (World Scientific, Singapore, 2000)

[5] Greffet, J.-J. Nanoantennas for Light Emission. *Science* **308**, 1561 (2005).

[6] Schuck, P. J., Fromm, D. P., Sundaramurthy, A., Kino, G. S. & Moerner, W. E. Improving the mismatch between light and nanoscale objects with gold bowtie nanoantennas. *Phys. Rev. Lett.* **94**, 017402 (2005).

[7] Mühlschlegel, P., Eisler, H.-J., Martin, O. J. F., Hecht, B. & Pohl, D. W. Resonant optical antennas. *Science* **308**, 1607 (2005).

[8] Takahara, J., Yamagishi, S., Taki, H., Morimoto, A. & Kobayashi, T. Guiding of a one-dimensional optical beam with nanometer diameter. *Opt. Lett.* **22**, 475 (1997).

[9] Bozhevolnyi, S. I., Volkov, V. S., Devaus, E., Laluet, J.-Y. & Ebbesen, T. W. Channel plasmon subwavelength waveguide components including interferometers and ring resonators. *Nature* **440**, 508 (2006).

[10] Chang, D. E., Sørensen, A. S., Demler, E. A. & Lukin, M. D. A single-photon transistor using nanoscale surface plasmons. *Nature Physics* **3**, 807 (2007).

[11] Sukharev, M. & Sideman, T. Coherent control approaches to light guidance in the nanoscale. *J. Chem Phys.* **124**, 144707 (2006).

[12] Aeschlimann, M., Bauer, M., Bayer, D., Brixner, T., García de Abajo, F. J., Pfeiffer, W., Rohmer, M., Spindler, C. & Steeb, F. Adaptive subwavelength control of nano-optical fields. *Nature*, **446**, 301 (2007).

[13] Dickson, W., Wurtz, G. A., Evans, P. R., Pollard, R. J. & Zayats, A. V. Electronically Controlled Surface Plasmon Dispersion and Optical Transmission through Metallic Hole Arrays Using Liquid Crystal. *Nano Lett.* **8** , 281 (2008).

[14] Smolyaninov, I. I., Davis, C. C. & Zayats, A. V. Light-controlled photon tunnelling, *Appl. Phys. Lett.* **81**, 3314 (2002).

[15] Feigenbaum, E. & Orenstein, M. Perfect 4-way splitting in nano plasmonic X-junctions. *Opt. Express* **15**, 17948 (2007).

[16] Pacifici, D., Lezec, H. J. & Atwater, H. A. All-optical modulation by plasmonic excitation of CdSe quantum dots. *Nature Photonics* **1**, 402 (2007).

[17] Maier, S. A., Kik, P. G., Atwater, H. A., Meltzer, S., Harel, E., Koel, B. E. & Requicha, A. A. G. Local detection of electromagnetic energy transport below the diffraction limit in metal nanoparticle plasmon waveguides. *Nature Materials* **2**, 229 (2003).

[18] Krenn, J. R., Dereux, A., Weeber, J. C., Bourillot, E., Lacroute, Y., Goudonnet, J. P., Schider, G., Gotschy, W., Leitner, A., Aussenegg, F. R. & Girard, C. Squeezing the Optical Near-Field Zone by Plasmon Coupling of Metallic Nanoparticles. *Phys. Rev. Lett.* **82**, 2590 (1999).

[19] Krenn, J. R. & Weber, Surface plasmon polaritons in metal stripes and wires. *J. C. Phil. Trans. R. Soc. Lond. A* **362**, 739 (2004).

[20] Ditlbacher, H., Hohenau, A., Wagner, D., Kreibig, U., Rogers, M., Hofer, F., Aussenegg, F. R. & Krenn, J. R. Silver Nanowires as surface plasmon resonators. *Phys. Rev. Lett.* **95**, 257403 (2005).

[21] Bozhevolnyi, S. I., Erland, J., Leosson, K., Skovgaard, P. M. W. & Hvam, J. M. Waveguiding in Surface Plasmon Polariton Band Gap Structures. *Phys. Rev. Lett.* **86**, 3008 (2001).

[22] Tanaka, K. & Tanaka, M. Simulations of nanometric optical circuits based on surface plasmon polariton gap waveguide, *Appl. Phys. Lett.* **82**, 1158 (2003).

[23] Liu, L., Han, H. & He, S. Novel surface plasmon waveguide for high Integration. *Opt. Express* **13**, 6645 (2005).

[24] Lee, I., Jung, J., Park, J., Kim, H. & Lee, B. Dispersion characteristics of channel plasmon polariton waveguides with step-trench-type grooves. *Opt. Express* **15**, 16596 (2007).

[25] Bozhevolnyi, S. I., Volkov, V. S., Devaux, E. & Ebbesen, T. W. Channel Plasmon-Polariton Guiding by Subwavelength Metal Grooves. *Phys. Rev. Lett.* **95**, 046802 (2005).

[26] Volkov, V. S., Bozhevolnyi, S. I., Devaux, E., Laluet, J.-Y. & Ebbesen, T. W. Wavelength Selective Nanophotonic Components Utilizing Channel Plasmon Polaritons. *Nano Lett.* **7**, 880, (2007).

[27] Farahani, J. N., Pohl, D. W., Eisler, H.-J. & Hecht, B. Single quantum dot coupled to a scanning optical antenna: a tunable superemitter. *Phys. Rev. Lett.* **95**, 017402 (2005).

[28] Kühn, S., Hakanson, U., Rogobete, L. & Sandoghdar, V. Enhancement of single molecule fluorescence using a gold nanoparticle as an optical nano-antenna. *Phys. Rev. Lett.* **97**, 017402 (2006).

[29] Kim, S., Jin, J., Kim, Y.-J., Park, I.-Y., Kim, Y. & Kim, S.-W. High-harmonic generation by resonant plasmon field enhancement. *Nature* **453**, 757 (2008).

[30] Ghenuche, P., Cherukulappurath, S., Taminiau, T. H., van Hulst, N. F. & Quidant, R. Spectroscopic Mode Mapping of Resonant Plasmon Nanoantennas. *Phys. Rev. Lett.* **101**, 116805 (2008).

[31] Anger, P., Bharadwaj, P. & Novotny, L. Enhancement and Quenching of Single-Molecule Fluorescence. *Phys. Rev. Lett.* **96**, 113002 (2006).

[32] Cheng, D. K. *Field and Wave Electromagnetics* (Addison Wesley, New York, 1983).

[33] Engheta, N., Salandrino, A. & Alù, A. Circuit elements at optical frequencies: nanoinductors, nanocapacitors, and nanoresistors. *Phys. Rev. Lett.* **95**, 095504 (2005).

[34] Alù, A. & Engheta, N. Tuning the scattering response of optical nanoantennas with nanocircuit loads. *Nature Photonics* **2**, 307 (2008).

[35] Johnson, P. B. & Christy, R. W. Optical constants of noble metals. *Phys. Rev. B* **6**, 4370 (1972).

[36] Novotny, L. Effective wavelength scaling for optical antennas. *Phys. Rev. Lett.* **98**, 266802 (2007).

[37] Barnard, E. S., White, J. S., Chandran, A. & Brongersma M. L. Spectral properties of plasmonic resonator antennas. *Opt. Express* **16**, 16529 (2008).

[38] Zhu, Z. & Brown, T. G. Full-vectorial finite-difference analysis of microstructured optical fibers. *Opt. Express* **10**, 853 (2002).

[39] Cinchetti, M., Gloskovskii, A., Nepjiko, S. A., Schonhense, G., Rochholz, H. & Kreiter, M. Photoemission Electron Microscopy as a Tool for the Investigation of Optical Near Fields. *Phys. Rev. Lett.* **95**, 047601 (2005).

[40] Douillard, L., Charra, F., Korczak, Z., Bachelot, R., Kostcheev, S., Lerondel, G., Adam, P.-M. & Royer, P. Short range plasmon resonators probed by photoemission electron microscopy. *Nano Lett.* **8**, 935 (2008).

[41] Weeber, J.-C., Krenn, J. R., Dereux, A., Lamprecht, B., Lacroute, Y. & Goudonnet, J. P. Near-field observation of surface plasmon polariton propagation on thin metal stripes. *Phys. Rev. B* **64**, 045411 (2001).

[42] Zentgraf, T., Dorfmüller, J., Rockstuhl, C., Etrich, C., Vogelgesang, R., Kern, K., Pertsch, T., Lederer, F. & Giessen, H. Amplitude- and phase-resolved optical near fields of split-ring-resonator-based metamaterials. *Opt. Lett.* **33**, 848 (2008).

[43] Lee, K. F. *Principles of antenna theory* (John Wiley & Sons, 1984).

[44] Rahola, J. Power Waves and Conjugate Matching. *IEEE Trans. Circuits and Systems II: Express briefs* **55**, 92 (2008).

[45] Kurokawa, K. Power waves and the scattering matrix. *IEEE Trans. Microw. Theory Tech.* **13**, 194 (1965).

[46] Nikitin, P. V., Rao, K. V. S., Lam, S. F., Pillai, V., Martinez, R. & Heinrich, H. Power Reflection Coefficient Analysis for Complex Impedances in RFID Tag Design. *IEEE Trans. Microw. Theory Tech.* **53**, 2721 (2005).

# Supplementary Material

# Impedance matching and emission properties of optical antennas in a nanophotonic circuit

Jer-Shing Huang, Thorsten Feichtner, Paolo Biagioni & Bert Hecht

**I. Determination of the propagation constants of TE guided modes on an infinitely long optical two-wire transmission line by nonlinear fitting.**

To obtain the propagation constants of guided TE modes in the optical transmission line (OTL) under consideration, we record the near-field intensity along the dielectric gap of the infinitely long OTL ($L$ = 3000 nm) and apply a non-linear fitting procedure. Due to a spurious reflection at the boundary of the simulation volume, we obtain a low amplitude standing wave pattern which we exploit to determine the wavelength of the fundamental mode (see Fig. 2B in the main manuscript). Furthermore, we observe longer-range undulations close to the receiving antenna due to a beating of several higher-order modes. In order to fit the simulation results with sufficient accuracy, we introduce three modes, $A_1$, $A_2$ and $A_3$, and apply a stepwise fitting procedure as described in the following. We define the voltage amplitude of the i$^{th}$ forward propagating mode at position $x$ in the OTL gap as $A_{0,i} e^{-\gamma_i x} e^{j\theta_i}$, where j is the imaginary unit, $A_{0,i}$ is the maximum field amplitude of the i$^{th}$ mode, and $\gamma_i = \alpha_i + j\beta_i$ its propagation constant consisting of the decay constant $\alpha_i$ and the wave vector $\beta_i$ of the mode. Finally, $\theta_i$ is the phase difference between the i$^{th}$ mode and the fundamental mode (i=1). Since the undulations cease after a distance of 500 nm (see Fig. 2B in the main manuscript), it is reasonable to assume that only the fundamental mode ($A_1$) reaches the boundary ($L$ = 3000 nm) with significant amplitude and is reflected (see Fig. S1 B). Moreover, due to the considerable damping of the gold OTL, we consider only a single reflection at the load and neglect contributions from all further reflections. Using the complex voltage reflection coefficient $\Gamma_V = |\Gamma_V|e^{j\theta_\Gamma}$ for the fundamental mode at the reflecting point, the overall field amplitude $A_i$ for each of the three modes can be described as

$$\begin{cases} A_1 = A_{0,1}e^{-\gamma_1 x}e^{j\theta_1}\left(1+\Gamma_V e^{-2\gamma_1(L-x)}\right) \\ A_2 = A_{0,2}e^{-\gamma_2 x}e^{j\theta_2} \\ A_3 = A_{0,3}e^{-\gamma_3 x}e^{j\theta_3} \end{cases} \quad (4)$$

and the total intensity $I$ is

$$I = \left|\sum_i A_i\right|^2. \quad (5)$$

In the first step, we fit the data from $x = 1000$ nm to $x = 3000$ nm, considering only the fundamental mode ($A_1$), and obtain $A_{0,1}$, $\alpha_1$, $\beta_1$, $|\Gamma_V|$, and $\theta_\Gamma$. In the second step, we introduce one additional mode ($A_2$) to fit the field intensity between $x = 500$ nm and $x = 3000$ nm. Note that all parameters obtained in the first step are now kept fixed and four new parameters ($A_{0,2}$, $\alpha_2$, $\beta_2$, and $\theta_2$) are allowed to vary. Finally, we introduce a third mode to fit the whole profile ($x = 0$ to 3000 nm) where again we keep all previously determined parameters fixed and only allow for variations of the parameters of the new mode ($A_{0,3}$, $\alpha_3$, $\beta_3$, and $\theta_3$).

The field intensity profile and the resulting best fit to the data are shown in Fig. S1 A, while the field amplitudes vs. position for the different modes in the OTL are plotted in Fig S1 B. Propagation constants of the modes obtained from the stepwise fitting procedure are $\gamma_1 = (0.00084+0.02802j)$ nm$^{-1}$, $\gamma_2 = (0.00093+0.01009j)$ nm$^{-1}$, and $\gamma_3 = (0.01002+0.05502j)$ nm$^{-1}$, and the voltage reflectivity at the boundary is $|\Gamma_V| = 36\%$. These results clearly show that the fundamental mode sufficiently decays in amplitude (down to $< 0.3\%\ A_0$) before reaching back to the receiving antenna, which supports our treating of the OTL as an infinitely long waveguide in this case. In addition, the effective wavelength of the fundamental mode, $\lambda_{\text{eff}} = 2\pi/\beta_1 = 224$ nm, agrees very well with results obtained from an independent finite-difference frequency-domain (FDFD) calculation, which yields $\lambda_{\text{eff}}$ = 223 nm. The propagation velocity of the fundamental mode on the transmission line is about 27% of the speed of light in vacuum, which is smaller than its radio-frequency counter part. The decay constant of the fundamental mode $\alpha_1$ also agrees with that obtained from FDTD.

**II. Determination of the reflectivity of a finite length OTL (1344 nm) terminated by an emitting optical antenna.**

Now we apply the obtained propagation constants to find the voltage reflection coefficients in a finite length OTL terminated by an antenna of specific length and width. As in the first section, since $A_{0,2}$ is very small and $\alpha_3$ is very large, we assume that only the fundamental mode reaches the emitting antenna with significant field intensity, which means that the change of the emitting antenna dimensions affects only $\Gamma_V$ and the amplitude $A_{0,1}$ of the fundamental mode. Within this approximation, we are able to fit the complicated intensity profile (see e.g. Fig. 3, upper panel, main manuscript) with only 3 open parameters, i.e. $A_{0,1}$, $|\Gamma_V|$ and $\theta_V$. A typical best fit to the data is shown in Fig. 3, upper panel, main manuscript.

**III. Characteristic impedance**

Fig. S2 shows the distributions of the relevant components of the electric (**E**) and magnetic (**H**) field, respectively. **E** and **H** are obtained from FDFD simulations. The characteristic impedance of the two-wire OTL is defined as $Z_0 = V/I$, where $V$ is the voltage between two wires and $I$ is the current on the OTL at a given position. We obtain $V$ as a line integral of the complex electric field **E** from one wire core to the other using $V = \int \mathbf{E} \cdot d\mathbf{s}$ and evaluate $I$ according to Ampère's law using $I = \oint \mathbf{H} \cdot d\mathbf{s}$, where the typical closed-loop integration path is replaced with a sufficiently long path approximating a linear path from $y = -\infty$ to $y = +\infty$. The integration paths are indicated by red-dashed lines in Fig. S2. The OTL characteristic impedance thus obtained is $Z_0 = (216-5.5j)\ \Omega$.

## IV. Smith chart for power wave and power reflection

Since the characteristic impedance of the OTL is complex, we employ the concept of power waves and Kurokawa's method to describe the power flow in the system. Using the concept of power waves, the power reflection coefficient $\Gamma_P$ can be expressed as $\Gamma_P = \left|\dfrac{Z_L - Z_0^*}{Z_L + Z_0}\right|^2$, where $Z_0 = R_0 + jX_0$ is the impedance seen by the emitting antenna (load), and $Z_L = R_L + jX_L$ is the load impedance. Plotting the normalized complex impedance, $\bar{Z} = R_L + j(X_L + X_0)/R_0$, provides the Smith chart for the power waves, as shown in Fig. S3. In analogy to conventional Smith charts for voltage waves, the power reflection coefficient $\Gamma_P$ corresponds to the distance from the center of the chart to one particular point.

**Figures and captions:**

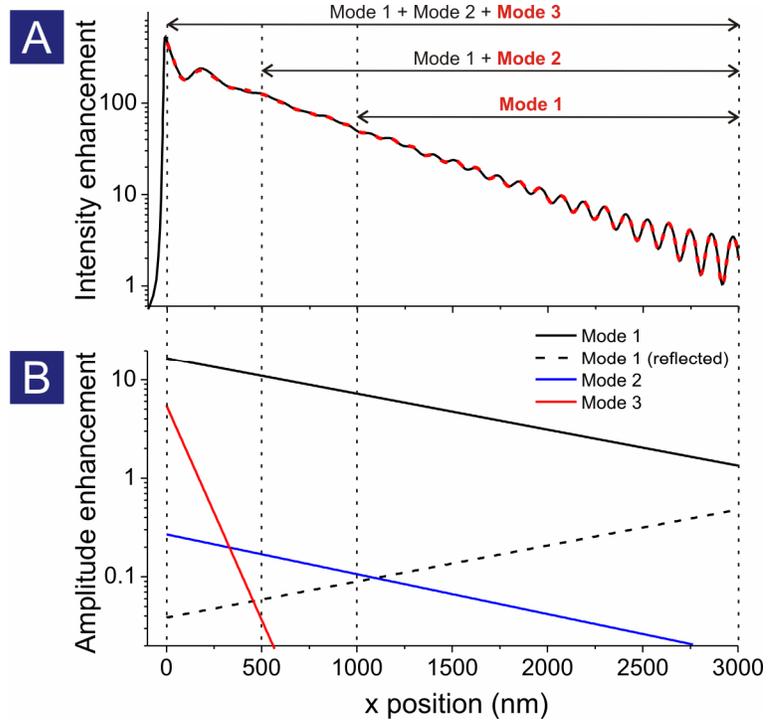

**Fig. S1** Modes on the infinitely long OTL. (A) The field intensity profile and the best fit (red dashed line) to the data obtained from a stepwise fitting. Data ranges and modes introduced in each fitting step are indicated above the double arrows. In each step, the mode whose parameters are allowed to vary is marked in red bold. (B) The field amplitudes as a function of position in the OTL of fundamental mode $A_1$ (black solid), its reflection (black dashed) and two higher-order modes ($A_2$ in blue and $A_3$ in red solid). Propagation constants of the modes obtained from the stepwise fitting procedure are $\gamma_1 = (0.00084+0.02802j)$ nm$^{-1}$, $\gamma_2 = (0.00093+0.01009j)$ nm$^{-1}$, and $\gamma_3=(0.01002+0.05502j)$ nm$^{-1}$. The voltage reflectivity for the fundamental mode at the boundary is $|\Gamma_V|=36\%$.

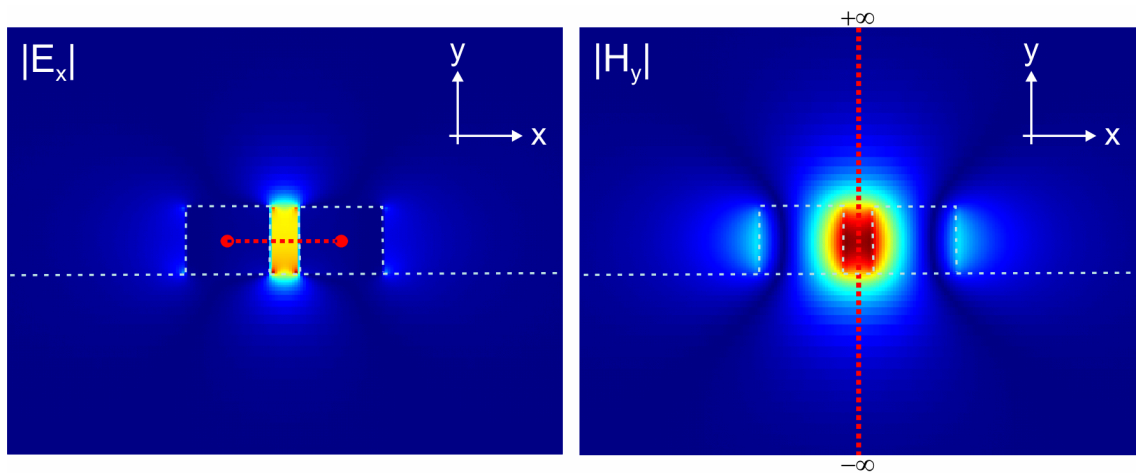

**Fig. S2** Distributions of the relevant electric (**E**) and magnetic (**H**) field components. The integration paths are indicated by the red-dashed lines. Since $H_y$ is zero at infinity, we replace the typical close loop integration path with an equivalent linear path and integrate $H_y$ from $y = -\infty$ to $y = +\infty$. The characteristic impedance of the two-wire OTL thus calculated is $Z_0 = V/I = (216 - 5.5j)\ \Omega$.

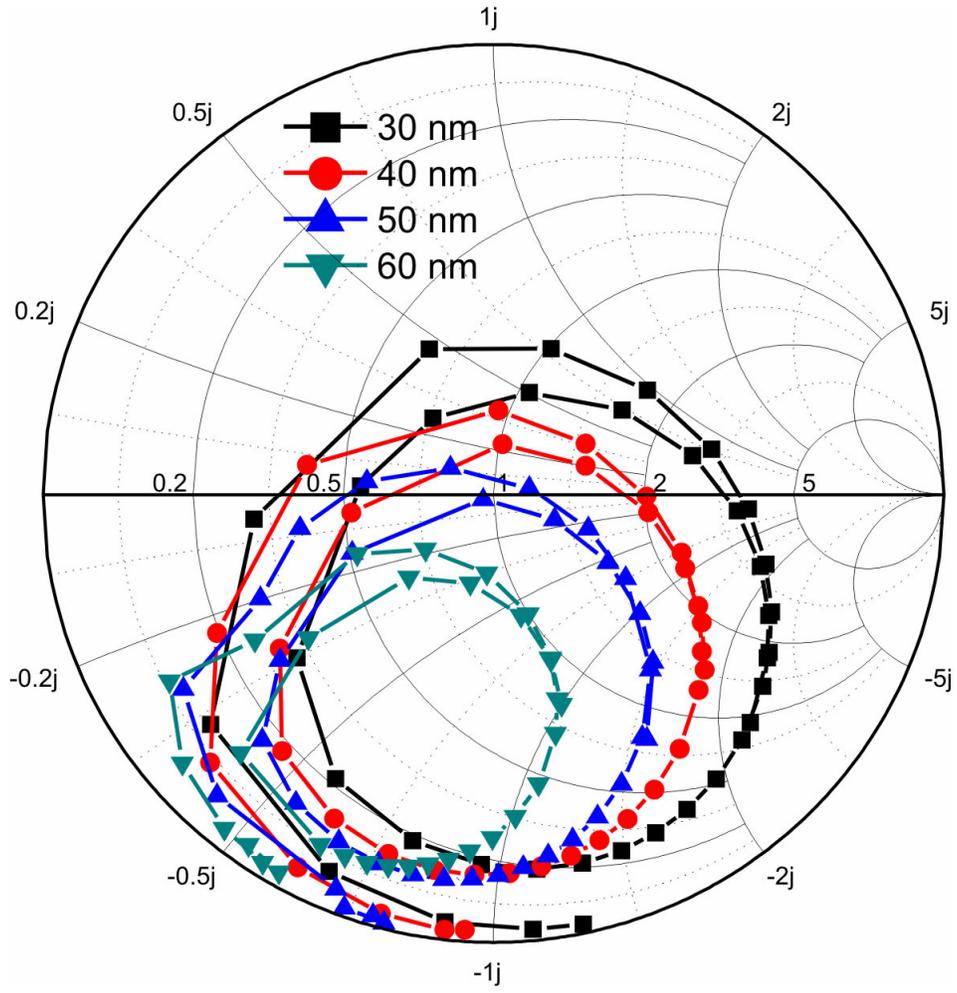

**Fig. S3** Smith chart mapping the normalized impedance $\bar{Z} = R_L + j(X_L + X_0)/R_0$. Distance from the center of the chart to one particular point corresponds to the power reflection coefficient $\Gamma_P = \left|\dfrac{Z_L - Z_0^*}{Z_L + Z_0}\right|^2$.